\documentclass{appolb}
\usepackage{amsmath}
\usepackage{amssymb}
\usepackage{amsfonts}
\usepackage{graphicx}
\usepackage{xcolor}
\usepackage{hyperref}
\usepackage{url}
\usepackage[sort&compress,numbers]{natbib}

\begin{document}

 \eqsec  
\title{Double pole $S$-matrix singularity in the continuum of $^7$Be%
\thanks{Presented at the XXXVIII Mazurian Lakes Conference on Physics, Piaski, Poland, August 31 --
September 6, 2025.}
}
\author{David Cardona Ochoa, Marek P{\l}oszajczak
\address{Grand Acc\'el\'erateur National d'Ions Lourds (GANIL), CEA/DSM - CNRS/IN2P3,
BP 55027, F-14076 Caen Cedex, France}
\\[3mm] 
Nicolas Michel
\address{CAS Key Laboratory of High Precision Nuclear Spectroscopy, Institute of Modern Physics, Chinese Academy of Sciences, Lanzhou 730000, China
\and
          School of Nuclear Science and Technology, University of Chinese Academy of Sciences, Beijing 100049, China}
          \\[3mm] 
Simin Wang
\address{Key Laboratory of Nuclear Physics and Ion-beam Application (MOE), Institute of Modern Physics, Fudan University, Shanghai 200433, China
\and
Shanghai Research Center for Theoretical Nuclear Physics, NSFC and Fudan University, Shanghai 200438, China}
}


\maketitle
\begin{abstract}
The double pole singularity of the $S$-matrix, the so-called exceptional point, associated with the  $5/2^-$ doublet of resonances in the spectrum of $^{7}$Be has been identified in the framework of the Gamow shell model. The exceptional point singularity is demonstrated by the coalescence of wave functions and spectral functions of the two resonances, as well as by the singular behavior of spectroscopic factors and electromagnetic transitions.
\end{abstract}
  
\section{Introduction}
\label{intro}

Exceptional points (EPs) are singularities that arise in quantum systems governed by non-Hermitian Hamiltonians~\cite{Ashida02072020}.
Unlike the Hermitian case, in which the crossing of energy levels with the same symmetry is avoided, EPs represent non-Hermitian degeneracies in the continuum, where two or more resonances and their corresponding eigenfunctions coalesce and the Hamiltonian becomes non-diagonalizable~\cite{PhysRevA.67.022721}. Near an EP, physical quantities often exhibit the non-analytic behavior and increased sensitivity to external perturbations \cite{Heiss_2012, PhysRevC.80.034619}.
 In this regime, eigenstates are strongly coupled through the environment, causing them to merge into a single inseparable mode and losing their individual identities \cite{Rotter_2015}.
 
Nuclear physics provides a natural setting for studying resonance phenomena. Resonances are a hallmark of systems that combine discrete quantum states to the continuum of scattering states through decays, captures, and virtual excitations. Moreover, the nuclear domain covers an exceptionally wide range of resonance lifetimes and widths, from narrow near-threshold states to broad resonances. This diversity enables systematic studies of resonance formation, evolution, and mixing, as well as the transition from bound states to the continuum. Nuclear physics thus provides both the conceptual groundwork and the experimental observables needed to test and refine theoretical descriptions of resonances, including non-Hermitian approaches and open quantum system frameworks such as the shell model embedded in the continuum~\cite{OKOLOWICZ2003271} which has been used before to study EPs~\cite{PhysRevC.80.034619}.

The Gamow shell model (GSM)~\cite{michel:in2p3-00331689, Michel2021} provides an open quantum system generalization of the standard shell model by incorporating, bound, resonant, and continuum single-particle states through a complex-energy Berggren basis, allowing to describe nuclear states in any regime of binding while preserving unitarity. It is thus a suitable tool for the study of exotic resonance phenomena such as EPs and to test their generic features and their specific effects on observables of particular interest in nuclear physics. 

\section{Theoretical Framework}\label{theory}

\subsection{Gamow shell model in coupled-channel representation}

The GSM extends the conventional nuclear shell model into the complex-energy plane, providing a framework for describing bound, resonant, and continuum states within the same formalism. In the GSM, A-body states are expressed as linear combinations of Slater determinants constructed from single-particle wave functions belonging to the Berggren ensemble~\cite{michel:in2p3-00331689, Michel2021}:  
\begin{equation}
    |\Psi^{J}_\alpha\rangle = \sum_n C_n |\text{SD}_n\rangle, \hspace{2mm} |\text{SD}_n\rangle = |\phi_{i_1} \phi_{i_2} \dots \phi_{i_A}\rangle,
\end{equation}  
where \( |\phi_{i}\rangle \) are the single-particle states, which include bound states, resonant (Gamow) states, and scattering states defined along a contour \(L_+\) in the complex momentum plane. The Berggren completeness relation ensures that the single-particle space is complete~\cite{michel:in2p3-00331689}:  
\begin{equation}
    \sum_{n \in \text{bound, res}} |\phi_n\rangle \langle \Tilde{\phi}_n| + \int_{L_+} |\phi(k)\rangle \langle \Tilde{\phi}(k)| \, dk = \hat{1}.
\end{equation}  
The many-body Hamiltonian $\hat{H}$ is then diagonalized in this complex-energy basis:  
   $ \hat{H} |\Psi^{J}_\alpha\rangle = E_\alpha |\Psi^{J}_\alpha\rangle$,
yielding complex eigenvalues $E_\alpha = \epsilon_\alpha - i  \Gamma_\alpha/2$, where the real and imaginary parts correspond to the energy and decay width of the nuclear state, respectively. This formalism allows the GSM to naturally describe the interplay between bound and continuum states, providing a microscopic and unified treatment of weakly bound nuclei, resonances, and the transition between discrete and continuum nuclear dynamics.

For a resonant state, with complex energy $E_\alpha = \epsilon_\alpha - i  \Gamma_\alpha/2$,  the time evolution of the wavefunction: 
\begin{equation}
|\Psi_\alpha(t)\rangle = e^{-i\epsilon_\alpha t/\hbar} e^{-\Gamma_\alpha t / 2\hbar} |\Psi_\alpha\rangle,
\end{equation}
exhibits an exponential decay with the halflife $T_{1/2} = \hbar\ln2/\Gamma_\alpha$. In this context, the expectation value of an operator $\hat{O}$ becomes complex:  
\begin{equation}
\label{expv}
\langle \hat{O} \rangle_\alpha = \frac{\langle \Tilde{\Psi}_\alpha | \hat{O} | \Psi_\alpha \rangle}{\langle \Tilde{\Psi}_\alpha | \Psi_\alpha \rangle} = \text{Re}\,\langle \hat{O} \rangle_\alpha + i\,\text{Im}\,\langle \hat{O} \rangle_\alpha.
\end{equation}
The real part corresponds to the mean value of the observable, as in the Hermitian case, while the imaginary part encodes the intrinsic dynamical uncertainty associated with the decay process, arising from the temporal instability of the resonant state~\cite{Moiseyev_2011}.

Although the Slater determinant formulation of the GSM provides a rigorous description of nuclear structure in open quantum systems, it is not well suited for calculating reaction observables because it treats the many-body states without explicitly distinguishing between different reaction channels and mass partitions. In this representation, scattering boundary conditions and the asymptotic behavior of reaction channels cannot be implemented, making it impossible to extract quantities such as cross sections or phase shifts.
The Gamow shell model in the coupled channels representation (GSM-CC)~\cite{Michel2021,PhysRevC.89.034624, Wang2017} is a convenient reformulation of the GSM that provides the necessary tools to study nuclear reaction and scattering properties. In this representation, the A-body wave function can be decomposed into the reaction channels:
\begin{equation}
    |\Psi^J\rangle = \sum_c \int_0^\infty |(c,r)^J\rangle \frac{u_c^{J}(r)}{r}r^2 dr ,
    \label{react_channels}
\end{equation}
where $r$ is the relative distance between the centers of mass of the projectile and target and where the projection $M$ of the total angular momentum $J$ is not explicitly written as results are independent of $M$. 
$u_c^{J}(r)$ is the radial amplitude describing the relative motion between the target and the projectile in the channel $c$ and is the solution to be determined by solving the GSM-CC equations for every given total angular momentum $J$. It should be emphasized that, in principle, the summation over channels should include all possible decay modes corresponding to binary, ternary, and more complicated mass partitions. In practice, however, the expansion is truncated, and so far applications of GSM-CC have been restricted mainly to binary channels (see however Ref. \cite{Wang2017,PhysRevC.108.044616}). The binary channel states $|(c,r)^J\rangle$ are defined as an antisymmetrized tensor product between the target state $|\Psi_T^{J_T}\rangle$ and the projectile state $|\Psi_P^{J_P}\rangle$ as follows:

\begin{equation}
 |(c,r)^J\rangle = \hat{\mathcal{A}}\left[ |\Psi_T^{J_T}\rangle \otimes |\Psi_P^{J_P}\rangle\right]^{J},
\end{equation}
where the channel index $c$ represents the mass partitions and their respective quantum numbers. The coupling between the angular momentum of the target $J_T$ and the projectile $J_P$ provides the total angular momentum $J$ of the system.
The coupled-channel equations can then be obtained from the Schr\"odinger equation $\hat{H}|\Psi^J\rangle = E |\Psi^J\rangle$, as:

\begin{equation}
    \sum_c \int_0^\infty r^2 \left(H_{cc'}(r, r') - E N_{cc'} (r,r')\right) \frac{u_c^J(r)}{r} =0, \label{schreq}
\end{equation}
where $E$ is the scattering energy of the full system, and the kernels are defined as:

\begin{equation}
    H_{cc'}(r,r') = \langle(c,r)^J|\hat{H}|(c',r')^J\rangle
\end{equation}
\begin{equation}
    N_{cc'}(r,r') = \langle(c,r)^J|(c',r')^J\rangle.
\end{equation}
As the nucleons of both clusters interact via short-range interactions, the Hamiltonian $\hat{H}$ can be written as:

\begin{equation}
    \hat{H}=\hat{H}_T + \hat{H}_P + \hat{H}_{TP},
\end{equation}
where $\hat{H}_T$ is the intrinsic Hamiltonian of the target and $\hat{H}_P$ is the Hamiltonian of the projectile which is decomposed as $\hat{H}_{P}=\hat{H}_{\rm int} + \hat{H}_{\rm CM}$, where $\hat{H}_{CM}$ describes the movement of the center of mass of the projectile and $\hat{H}_{int}$ is its intrinsic Hamiltonian. Additionally, $\hat{H}_{TP}$ is the Hamiltonian that represents the interaction between clusters defined as: $\hat{H}_{TP} = \hat{H}-\hat{H}_{T}-\hat{H}_{P}$, in which $\hat{H}$ is the standard shell model Hamiltonian.

\subsection{Exceptional points}

An EP occurs when for a Hamiltonian $H(\lambda)$  represented by a complex symmetric matrix depending on some set of parameters $\lambda$, one can find $\lambda = \lambda_{EP}$ such that at least two eigenvalues coalesce
\begin{equation}
    E_1(\lambda_{EP}) = E_2(\lambda_{EP}) = E_{EP},
\end{equation}
and their corresponding eigenfunctions become identical,
\begin{equation}
|\psi_1(\lambda_{EP})\rangle = |\psi_2(\lambda_{EP})\rangle = |\psi_{EP}\rangle,
\end{equation}
making Hamiltonian non-diagonalizable. 

In a non-Hermitian setting, the eigenpairs of $H(\lambda)$ satisfy:
\begin{equation}
    H(\lambda)|\psi_n(\lambda)\rangle = E_n(\lambda)|\psi_n(\lambda)\rangle, \hspace{2mm}  \langle \Tilde{\psi}_n(\lambda)| H(\lambda) =  \langle \Tilde{\psi}_n(\lambda)| E_n(\lambda),
\end{equation}
where $|\psi_n\rangle$ and $\langle \Tilde{\psi}_n|$ are the left and right eigenvectors respectively and $E_n$ is in general complex. 
In this setting, for ordinary eigenvectors which are not \textit{self-orthogonal}, one uses the biorthogonal normalization~  \cite{OKOLOWICZ2003271}: 
\begin{equation}
    \langle \tilde{\psi}_i | \psi_j \rangle = \delta_{ij}.
\end{equation}
which replaces the standard Hermitian scalar product. This modified normalization reflects the fact that left and right eigenvectors form a dual basis, and that their overlap carries physical information about the non-Hermitian character of the system.
A convenient measure of self-orthogonality is the \textit{phase rigidity}, defined as:
\begin{equation}
    r_i \equiv \frac{\langle \tilde{\psi}_i|\psi_i\rangle}{\langle \psi_i|\psi_i\rangle},
\end{equation}
which quantifies the degree of biorthogonality of the wave functions. For isolated, well-separated resonances, $r_i\approx1$, corresponding to a nearly Hermitian behavior. Approaching the EP, 
the states lose their separability and become self-orthogonal mixed through their mutual coupling to the continuum.
This self-orthogonality of the eigenfunctions at an exceptional point also has direct consequences for the expectation values of operators that do not commute with the Hamiltonian, $[\hat{H}, \hat{O}] \neq 0$. In the biorthogonal formalism, the expectation value of an operator $\hat{O}$ is given by Eq.~(\ref{expv}). As the exceptional point is approached, the eigenfunctions coalesce and become self-orthogonal, $\langle \tilde{\psi}_i(\lambda) | \psi_i(\lambda) \rangle  \to 0$ as $\lambda \to \lambda_{EP}$, causing the expectation value to diverge:
\begin{equation}
    \langle \hat{O} \rangle_i \to \infty \ .
    \label{equa}
\end{equation}
This divergence indicates extreme sensitivity to perturbations, non-analytic parameter dependence, and gives another signature of the loss of independence between the coalescing states.

\section{Results}\label{results}

We will discuss the effects of an exceptional point on quantities associated with the $5/2^-$ doublet of resonances present in the spectrum of $^7\text{Be}$. The nucleus in question was modeled by $^4\text{He}$ core with 3 valence nucleons and using a channel basis comprising $[^4\text{He}(0_1^+) \otimes {}^3\text{He}(L_j)]^{J^\pi}$, $[^5\text{Li}(K_i^\pi) \otimes {}^2\text{H}(L_j)]^{J^\pi}$ and $[^6\text{Li}(K_i^\pi) \otimes p(l_j)]^{J^\pi}$ channels. The cluster channels were constructed by coupling the partial waves $L_j = S_{1/2}$, $P_{1/2}$, $P_{3/2}$, $D_{3/2}$, $D_{5/2}$, $F_{5/2}$, $F_{7/2}$ of the wave function of ${}^3\text{He}$ with the inert core $^4\text{He}$ in the ground state $0^+$, and coupling the partial waves  $L_j $= ${^3S}_{1}$, ${^3P}_{0}$, ${^3P}_{1}$, ${^3P}_{3}$, ${^3D}_{1}$, ${^3D}_{2}$, ${^3D}_{3}$ of the wave function of $^2\text{H}$ with the $^5\text{Li}$ target in the ground state $3/2^-$. The one-proton channels were built by coupling the partial waves $l_j = s_{1/2}$, $p_{1/2}$, $p_{3/2}$, $d_{3/2}$, $d_{5/2}$, $f_{5/2}$, $f_{7/2}$ of the proton wave functions with the states $K_i^\pi= 1_1^+$, $3_1^+$, $0_1^+$ $2_1^+$, $2_2^+$, $1_2^+$ of $^6\text{Li}$.

\begin{figure}[h!tb]
\centering
\includegraphics[width=10cm,clip]{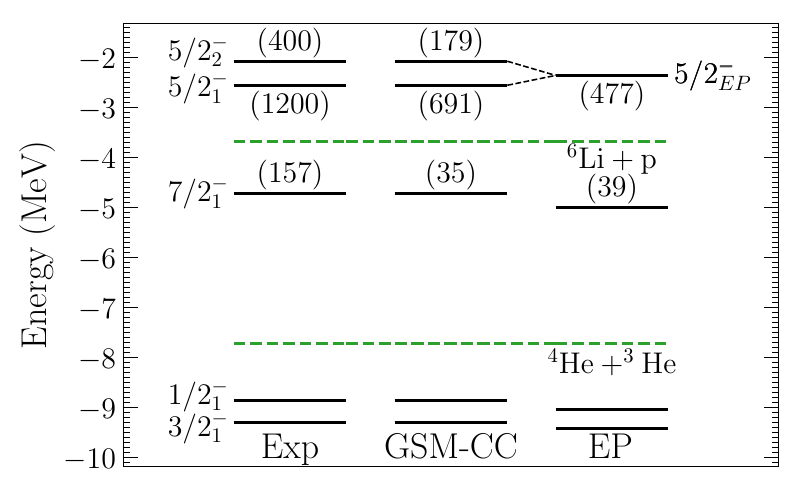}
\caption{The experimental spectrum of $^7\text{Be}$ is compared with the GSM-CC spectrum.  The widths (in keV) of resonances are given in brackets.  On the right-hand side, one shows the spectrum for a Hamiltonian $H(\lambda_{EP})$ for which the two $5/2^-$ states coalesce.}
\label{spec}       
\end{figure}

The Hamiltonian consists of a one-body part of the Woods-Saxon (WS) type plus a spin-orbit term and a Coulomb field mimicking the core, and a nucleon-nucleon interaction of the FHT type~\cite{10.1143/PTP.62.981}. Parameters of this Hamiltonian for $^7$Be have been adjusted~\cite{PhysRevC.108.044616} according to the data given in the ENSDF database~\cite{ENSDF}.
To account for the missing channels in Eq.~(\ref{react_channels}), the matrix elements for the the channel-channel couplings involving one-nucleon reaction channels have been re-scaled by corrective factor $c(J^\pi):$ $c(3/2^-)= 1.0008$,  $c(1/2^-)=1.007$, $c(7/2^-)=1.009$, $c(5/2^-)=0.9713$. Similarly, for the matrix elements involving cluster reaction channels, we have $c_{{}^3\text{He}}(5/2^-)=1.0152$ and  $c_{d}(5/2^-)=1.0945$ for ${}^3\text{He}$ and deuteron, respectively. 
Given this setup, Figure \ref{spec} shows the obtained energy spectrum relative to the energy of the $^4\text{He}$ core, along with the relevant particle emission thresholds. The calculated energy spectrum is in good agreement with the experimental data, yet the widths are slightly smaller than those given in ENSDF~\cite{ENSDF}. 
We used the strengths of $l=1$ spin-orbit one-body potential for protons and neutrons, $V^{(l=1)}_{SO}(p)$ and $V^{l=1)}_{SO}(n)$,  as the variable parameters of the Hamiltonian $H(\lambda)$ to find the EP in the position shown in Fig. \ref{spec}.

\begin{figure*}[h!tb]
\centering
\includegraphics[width=6.2cm,clip]{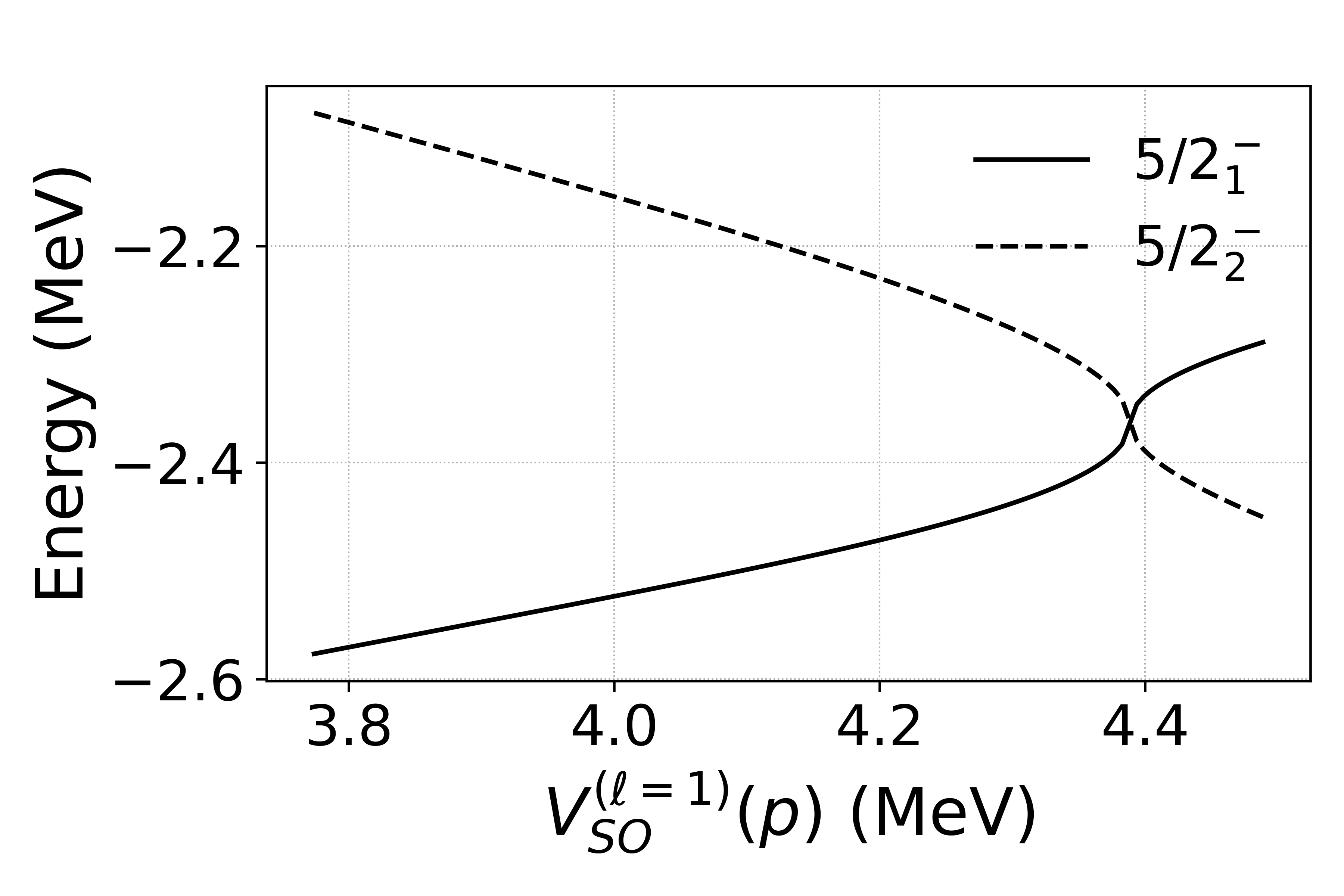}
\includegraphics[width=6.2cm,clip]{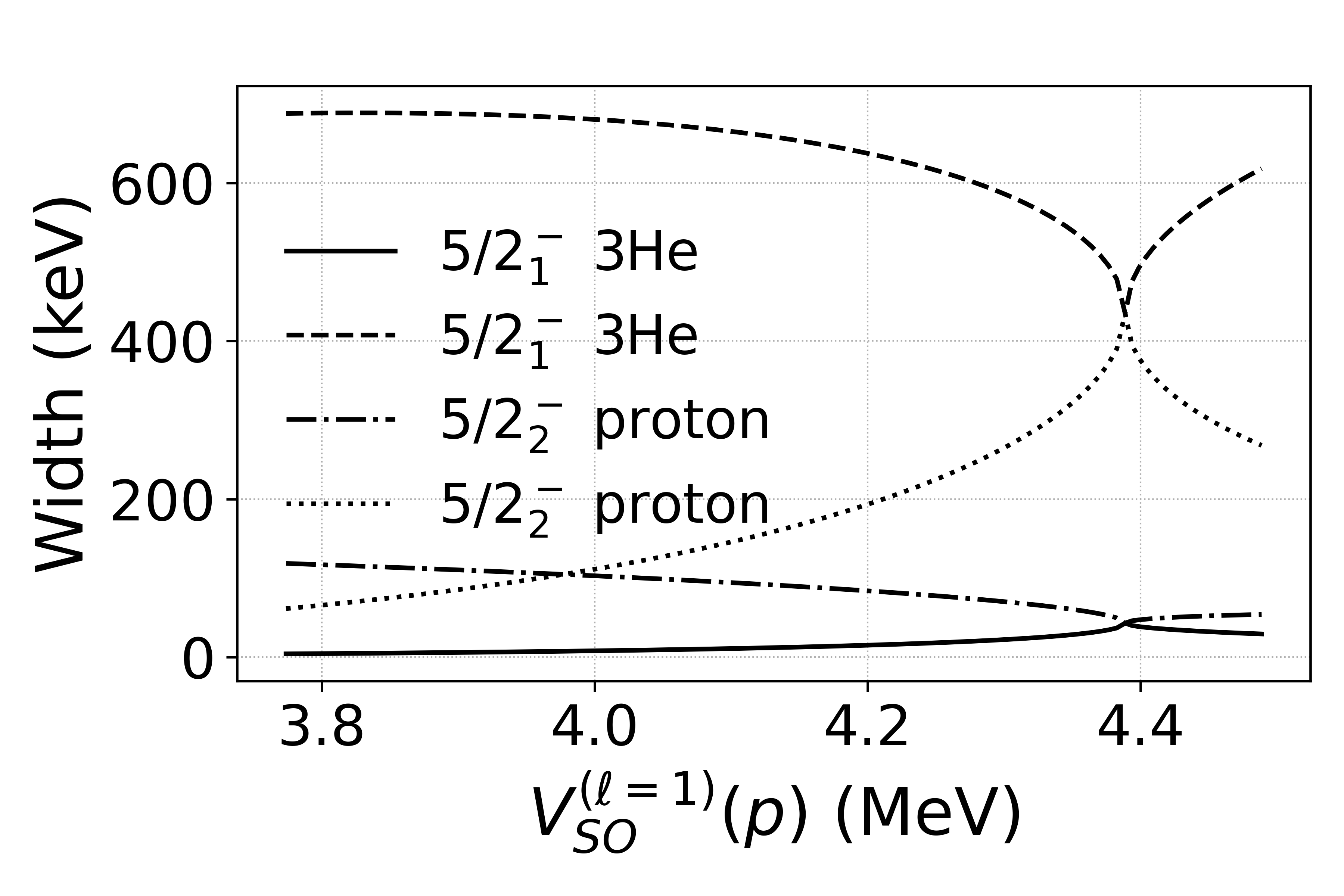}
\caption{The energy eigenvalue trajectories (left panel) and the partial widths of the proton and $^3$He resonances (right panel) are shown as a function of the spin-orbit potential strength $l=1$ for protons. Near EP, a characteristic square root dependence of the energy and width can be observed.
}
\label{ew}       
\end{figure*}

Figure \ref{ew} shows the coalescence of the energies and partial widths of the $5/2^-$ doublet as a function of the control parameters, starting from the fitted values $5/2^-_1:$ $E_1 = -2.557 \text{ MeV}$, $\Gamma_1=691 \text{ keV} $, and $5/2^-_2:$ $E_2 = -2.077 \text{ MeV}$, $\Gamma_2= 179 \text{ keV} $, and converging towards the found EP $5/2^-_{EP}:$ $E_{EP}= -2.36\text{ MeV}$, $\Gamma_{EP}=477\text{ keV}$, at $V^{(l=1)}_{SO}(p)= 4.38 \text{ MeV}$ and $V^{(l=1)}_{SO}(n)= 4.48 \text{ MeV}$. The coalescence of the partial decay widths indicates that the resonance wave functions  become identical. As the system approaches the EP, the analytic structure of the eigenvalues acquires a square-root singularity \cite{Heiss_2012}. This leads to intertwined trajectories of energies and widths that no longer vary independently.

\begin{figure*}[h!tb]
\centering
\includegraphics[width=10cm,clip]{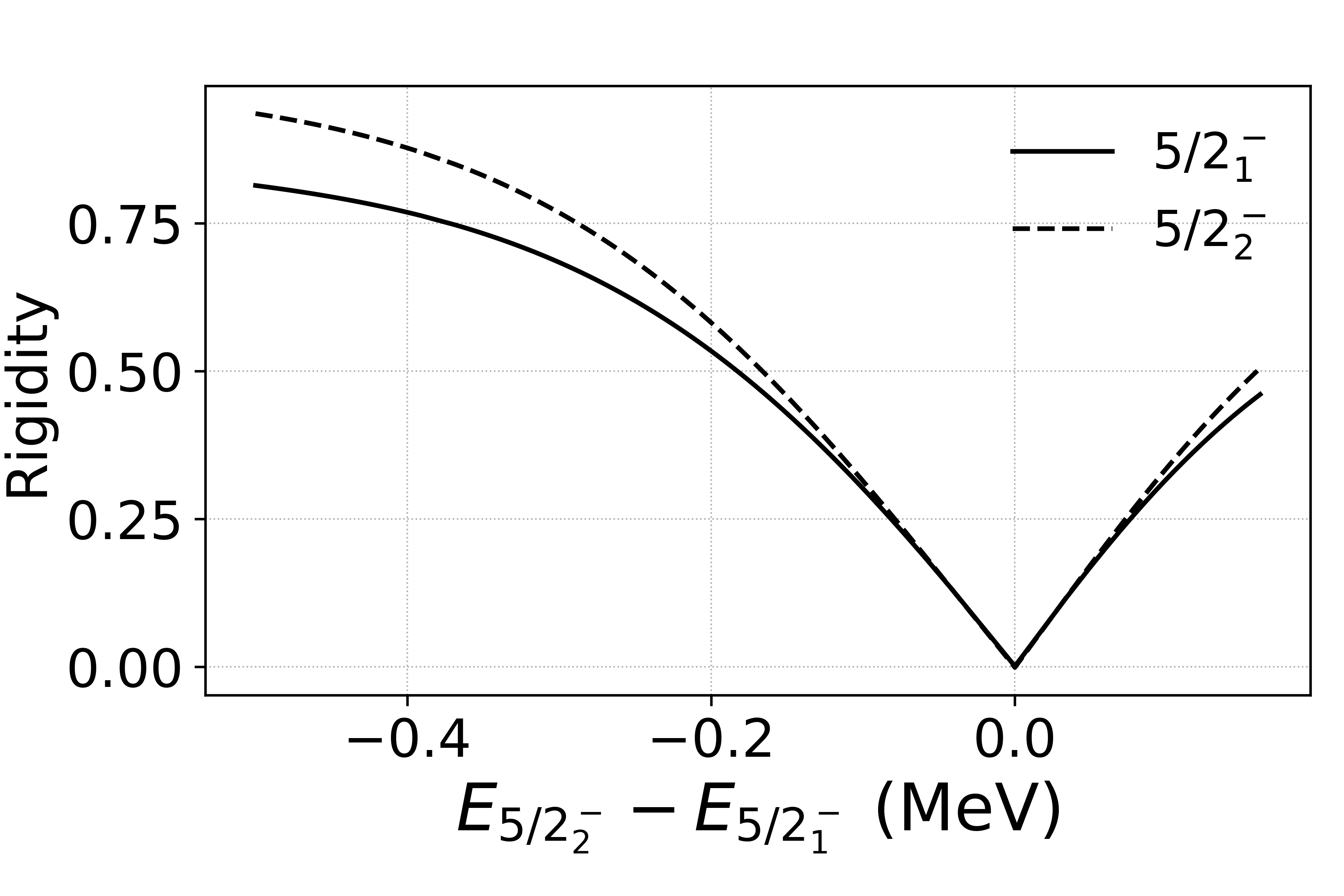}
\caption{The phase rigidity of $5/2^-$ resonances is plotted as a function of their separation energy. The phase rigidity becomes equal zero at the EP. }
\label{rig}       
\end{figure*}
Figure \ref{rig} illustrates the evolution of the phase rigidity for both $5/2^-$ states as a function of their energy distance. The initial departure from unity for both resonances indicates that the strong non-Hermitian effects are present already at their experimental energies. As the states approach one another, the rigidity tends to zero, reflecting their growing non-orthogonality and the strengthening of their mutual mixing through the coupling to common decay channels. 

\begin{figure*}[h!tb]
\centering
\includegraphics[width=6.2cm,clip]{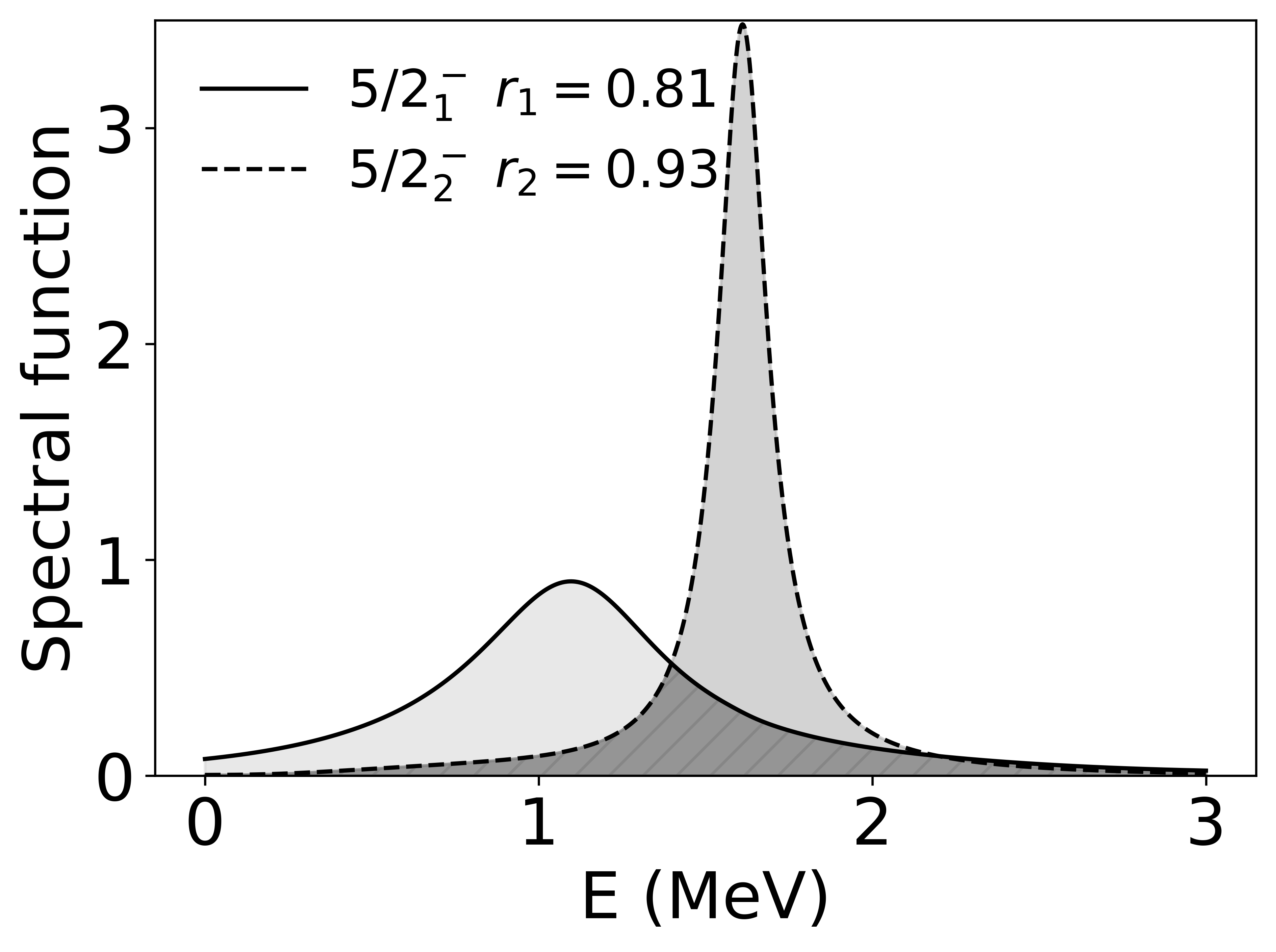}
\includegraphics[width=6.2cm,clip]{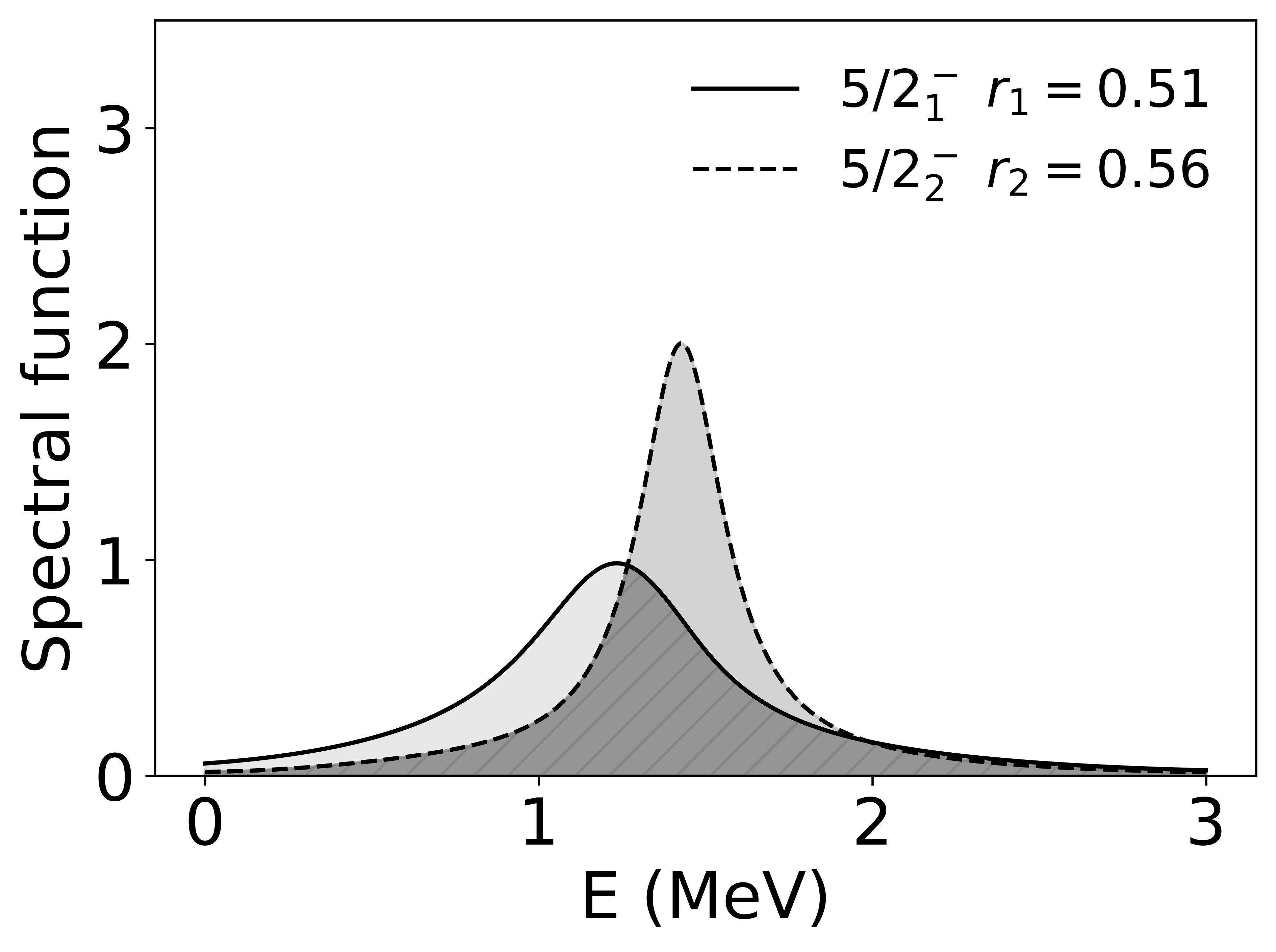}
\includegraphics[width=6.2cm,clip]{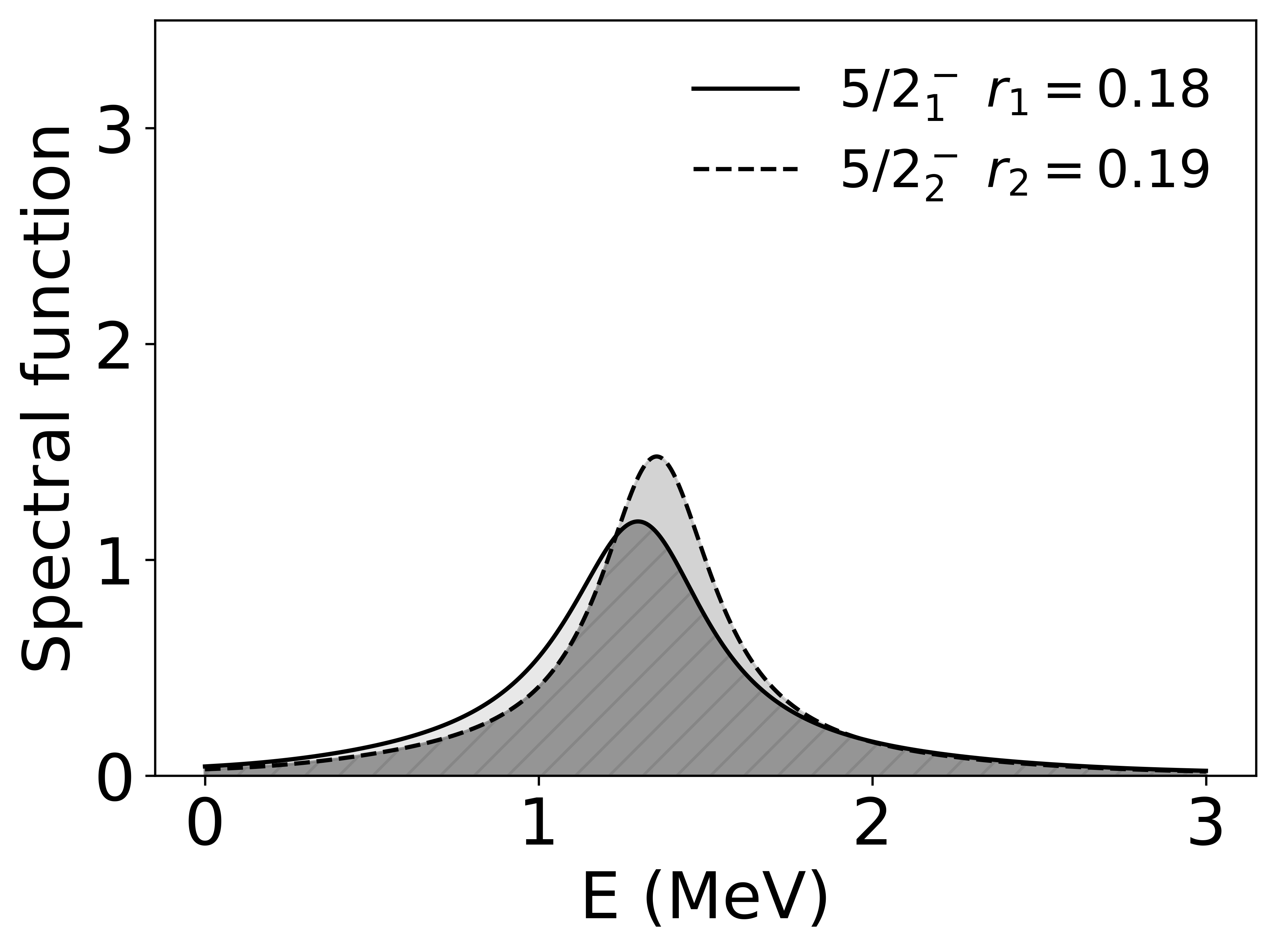}
\includegraphics[width=6.2cm,clip]{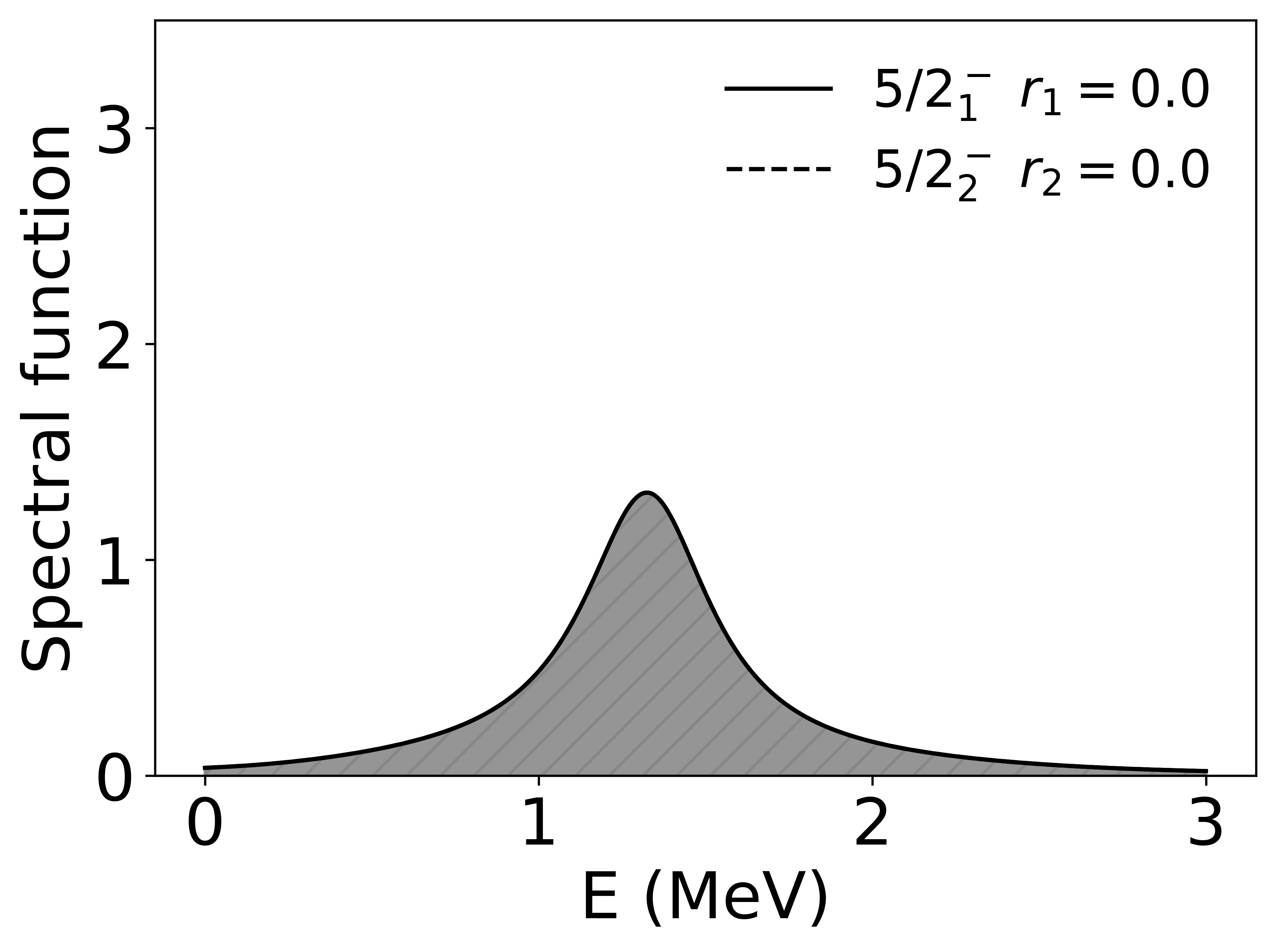}

\caption{Spectral function of both $5/2^-$ resonances for different values of the phase rigidity.}
\label{sp}       
\end{figure*}
Another quantity that provides insight on the coalescence of the eigenstates is the \textit{spectral function}. The spectral function $\widetilde{\rho}(E)$ can be expressed as the Fourier transform of the survival probability amplitude $\rho(t)$ of a many-body state $|\Psi\rangle$ : 
\begin{eqnarray}
\rho(t) &=& \mathcal{N}_\rho \sum_c \int_0^{+\infty} {u_c^J(r,0)}^* u_c^J(r,t)~dr \label{rho_t}, \\
\widetilde{\rho}(E) &=& \frac{1}{2 \pi} \int_{-\infty}^{+\infty} \rho(t) e^{i E t}~dt \label{rho_E},
\end{eqnarray}
where $u_c^J(r,t=0)$ is, in principle, arbitrary, and, in practice, is taken equal to $u_c^J(r)$ of Eq.(\ref{schreq}) multiplied by a smooth Fermi function slowly vanishing at infinity,
while $\rho(t)$ is normalized so that $\rho(t=0) = 1$, which fixes $\mathcal{N}_\rho$.

 $\widetilde{\rho}(E) $ characterizes the energy profile of the resonance, revealing its localization and decay characteristics. Figure \ref{sp} shows the spectral functions calculated at different values of rigidity, starting at the experimental fitted values. At the beginning of the evolution, the spectral functions of the two resonant states already show some degree of overlap, although their shapes and widths remain clearly distinct. As the rigidity decreases, these distributions gradually become more similar, reflecting the increasing mixing of the two states. At the EP, the spectral functions perfectly coincide, indicating that the two resonances are identical and have lost their individuality.

\begin{figure*}[h!tb]
\centering

\includegraphics[width=6.2cm,clip]{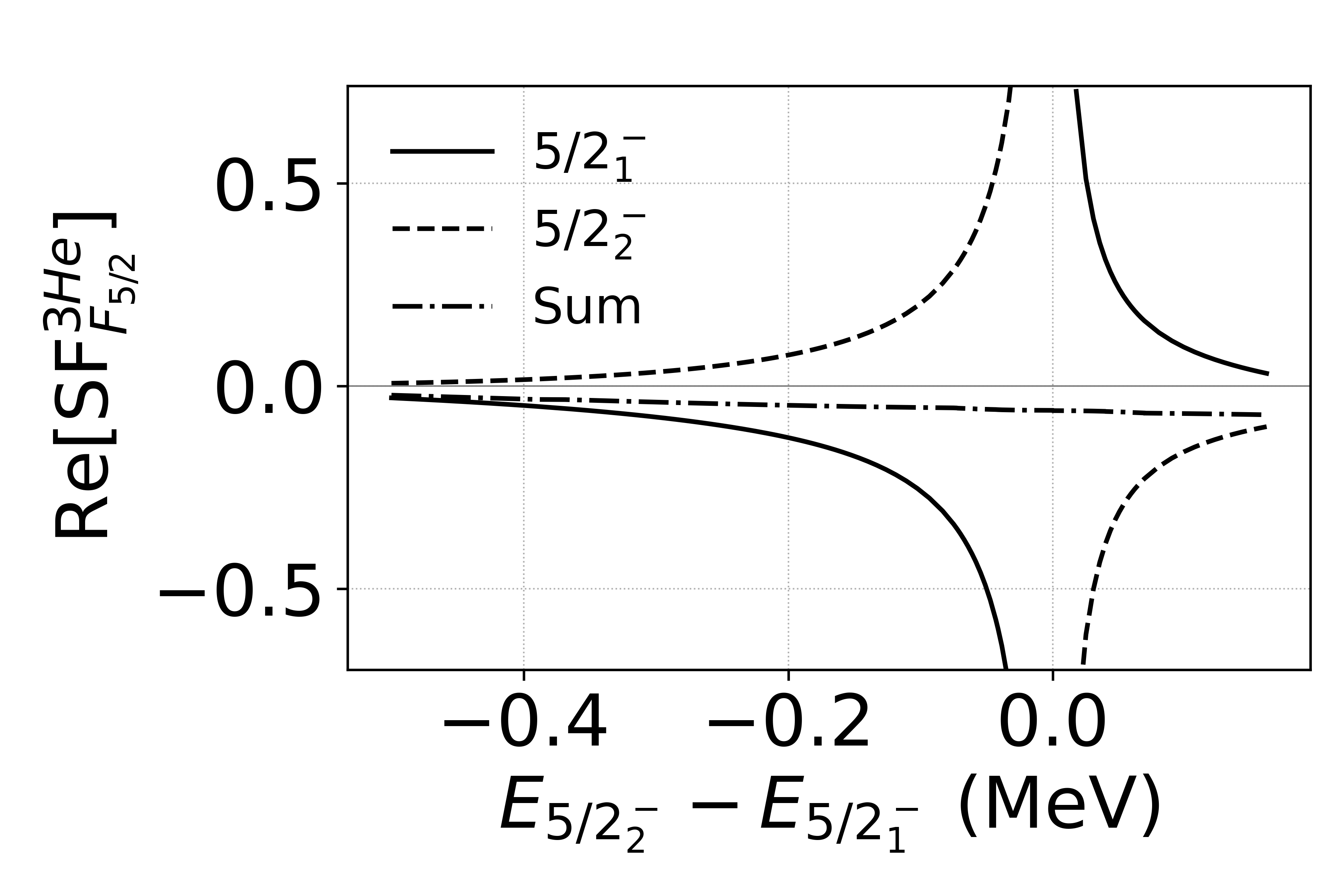}
\includegraphics[width=6.2cm,clip]{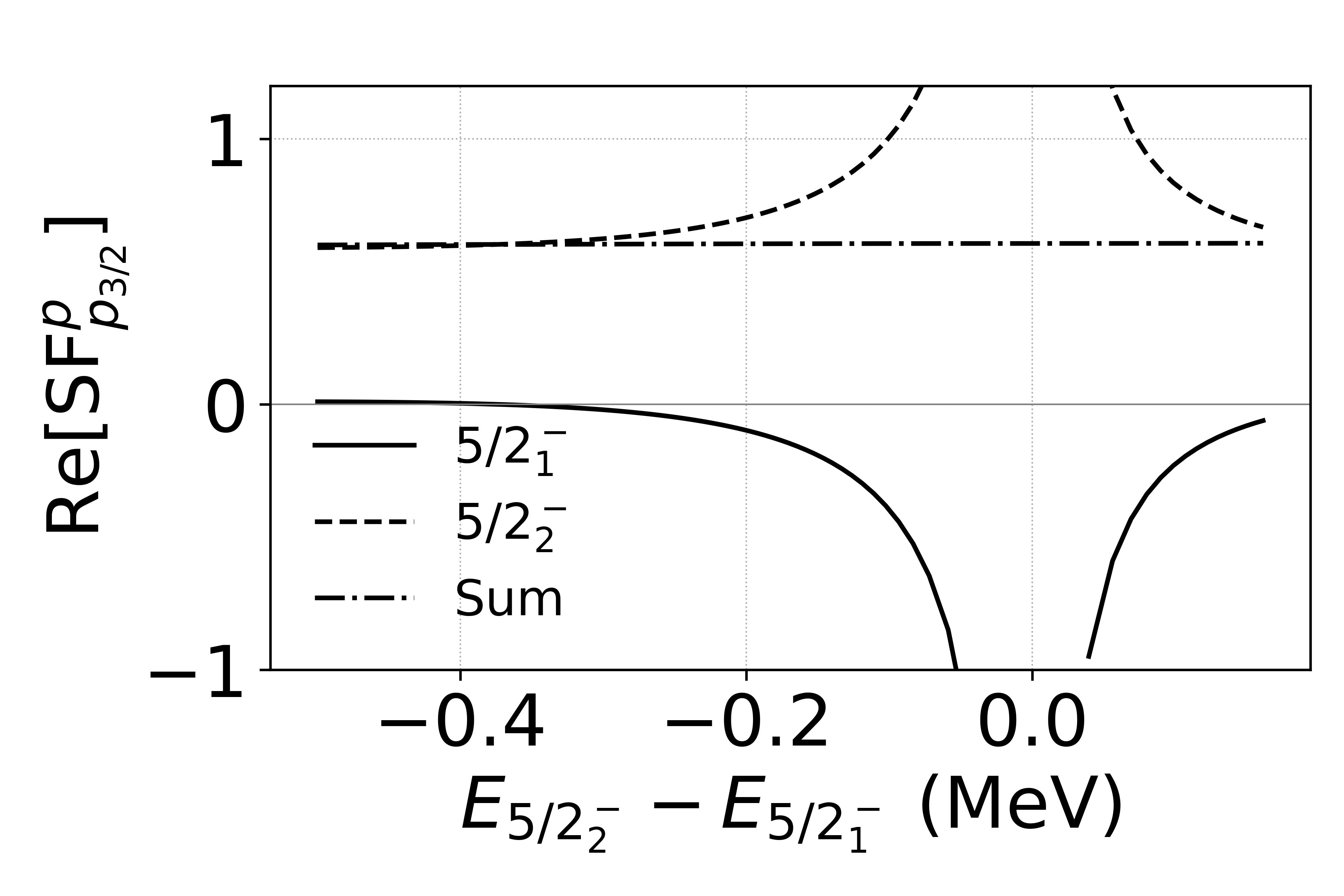}
\includegraphics[width=6.2cm,clip]{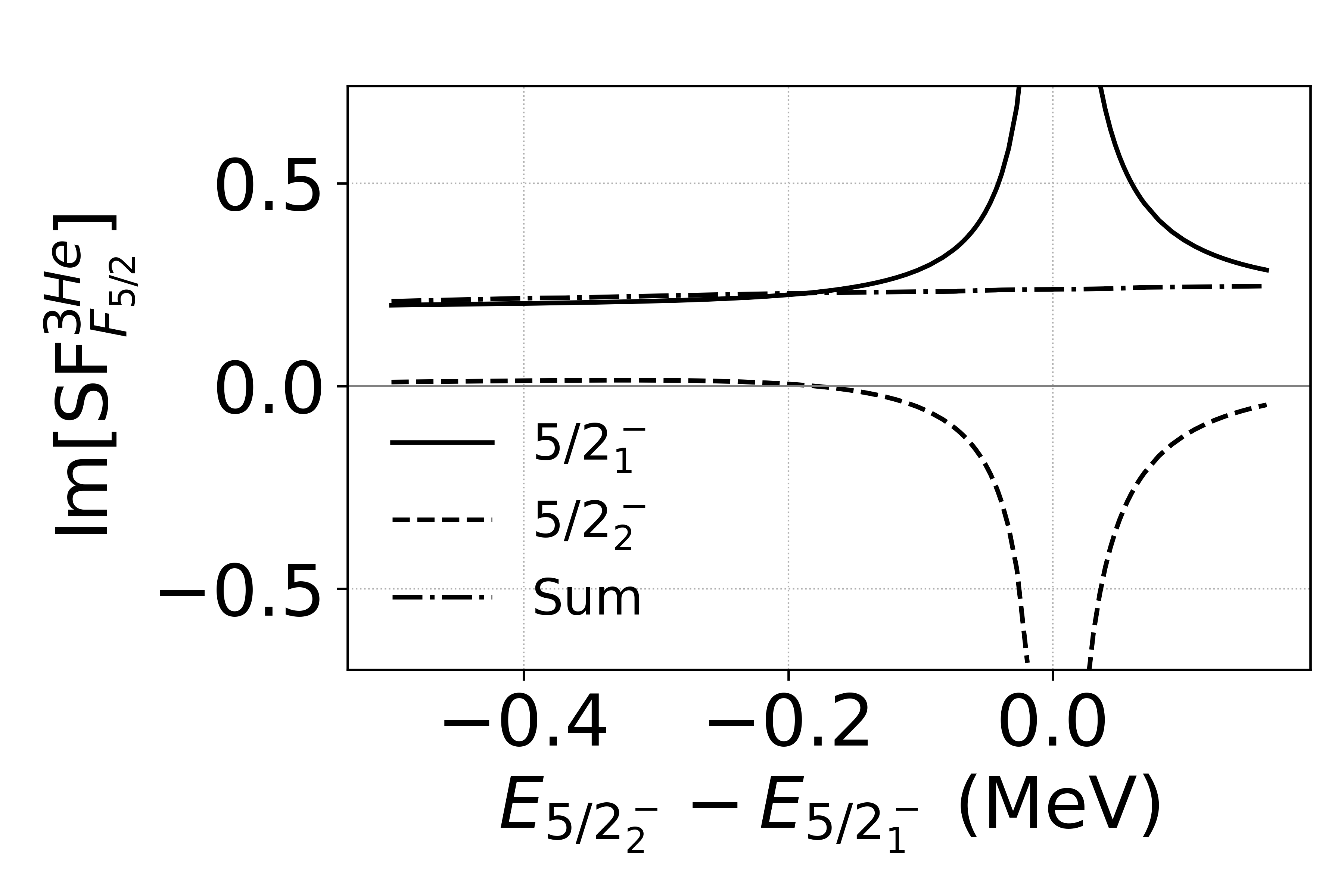}
\includegraphics[width=6.2cm,clip]{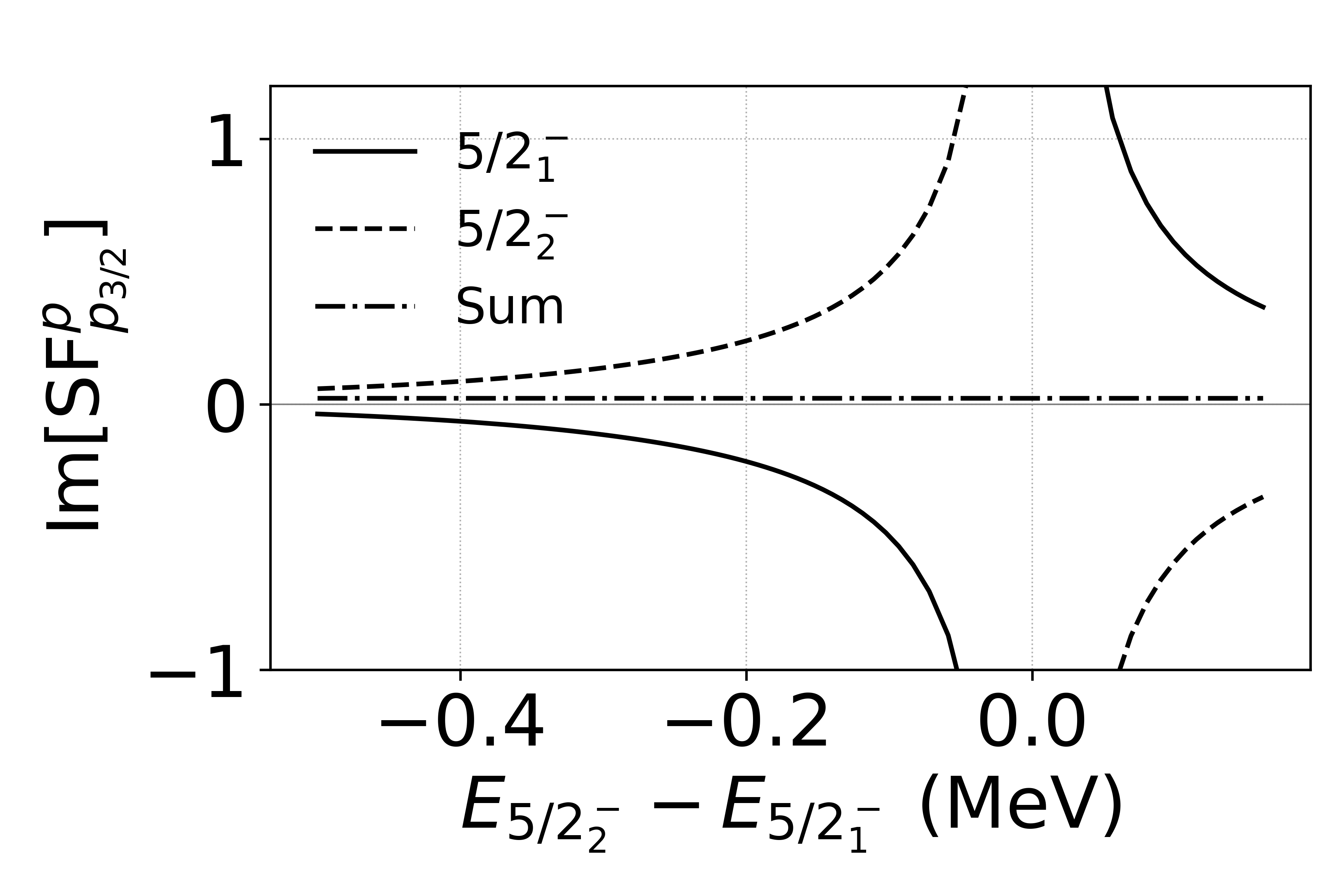}
\caption{Evolution of the real part (upper panels) and imaginary part (bottom panels) of the $\langle {}^7 \text{Be}(5/2^-_n) ||{}^4 \text{He}(0^+)\otimes{}^3 \text{He}(F_{5/2})\rangle$ (left panels) and $\langle {}^7 \text{Be}(5/2^-_n) ||{}^6 \text{Li}(1^+)\otimes\text{p}(p_{3/2})\rangle$  (right panels) spectroscopic factors for the $5/2^-_n$ ($n=1,2$) doublet of resonances in ${}^7 \text{Be}$ when approaching the EP as a function of the energy separation between these states.}
\label{Sf}       
\end{figure*}
\begin{figure*}[h!tb]
\centering
\includegraphics[width=6.2cm,clip]{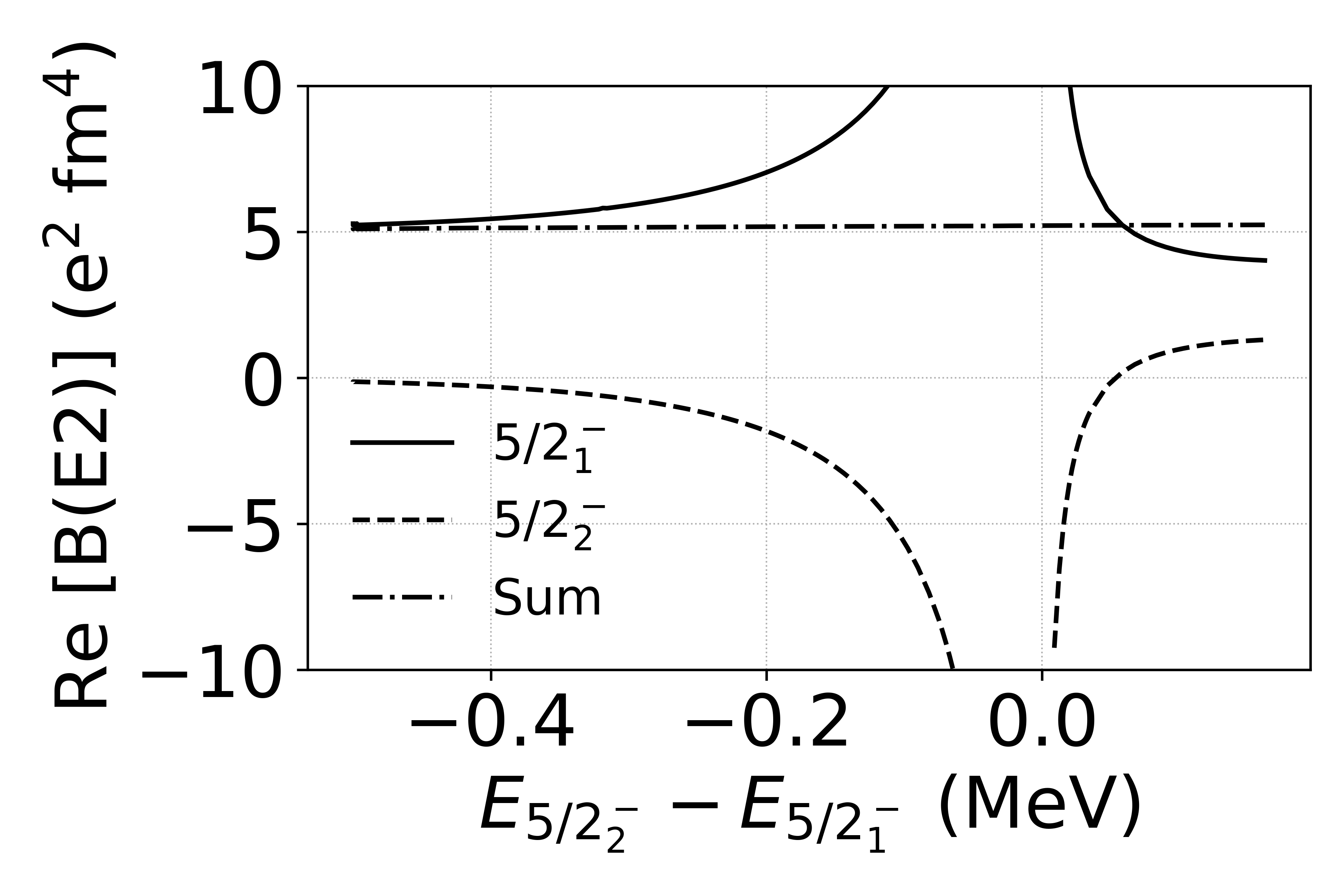}
\includegraphics[width=6.2cm,clip]{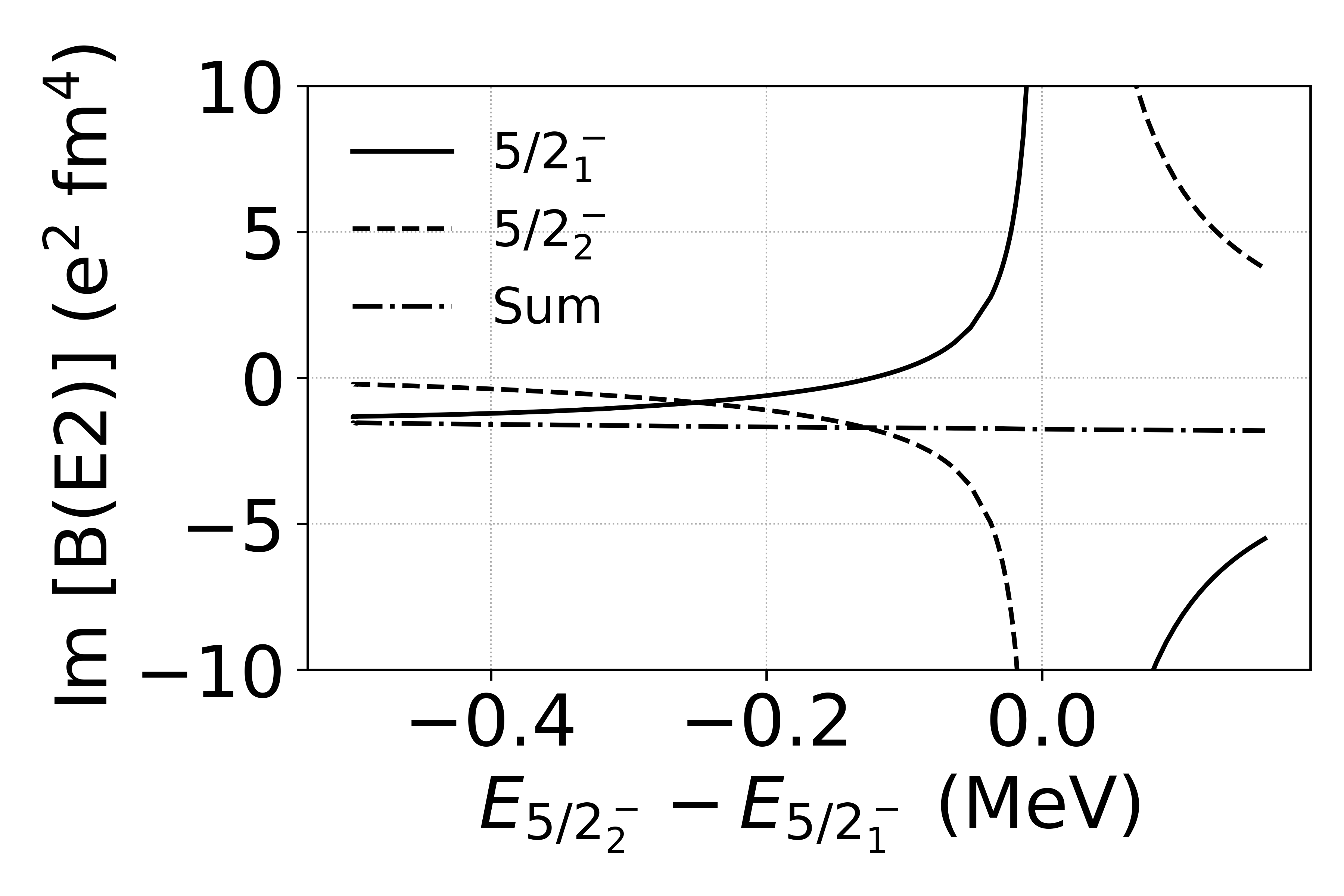}
\caption{Evolution of the real part (left panel) and imaginary part (right panel) of the $B(E2)$ transitions from the $5/2^-$ states to the $3/2^-$ ground state in ${}^7 \text{Be}$ when approaching the EP as a function of the energy separation between the states.}
\label{Bs}       
\end{figure*}

The more striking effects of the self-orthogonality can be seen in Fig. \ref{Sf}, where the evolution of the spectroscopic factors for $\langle {}^7 \text{Be}(5/2^-_n) ||{}^4 \text{He}(0^+)\otimes{}^3 \text{He}(F_{5/2})\rangle$ and $\langle {}^7 \text{Be}(5/2^-_n) ||{}^6 \text{Li}(1^+)\otimes\text{p}(p_{3/2})\rangle$ are shown as a function of the separation between the two states. Similarly, in Fig. \ref{Bs} we see an evolution of the $B(E2; 5/2^- \rightarrow 3/2^-)$ transition probabilities when approaching the EP. 
In GSM-CC, spectroscopic factors are calculated with the channel decomposition of $|\Psi\rangle$ from Eq.(\ref{react_channels}) : 
\begin{equation}
\langle \Psi_{A+a}^J ||~[\Psi_A^{J_A} \otimes{} \Psi_a^{J_a}(\ell j)]^J \rangle = \sqrt{\int_0^{+\infty} (u_{c_{SF}}^J(r))^2~dr} \label{SF},
\end{equation}
where $J$, $J_A$ and $J_a$ are, respectively, the total angular momentum of the composite wave function $|\Psi_{A+a}^J\rangle$ of $A+a$ nucleons,  of the target wave function, and of the projectile wave function $|\Psi_a^{J_a}(\ell j)\rangle$ of $a$ nucleons,  $\ell$ and $j$ are the orbital and total angular momenta of the center of mass of the projectile wave function $|\Psi_a^{J_a}(\ell j)\rangle$, respectively, while $c_{SF}$ is the channel whose target and projectile wave functions are $|\Psi_A^J\rangle$ and $|\Psi_a^{J_a}(\ell j)\rangle$, respectively. The integral over $r$ in Eq.(\ref{SF}) is calculated with complex scaling \cite{Michel2021}.

At the beginning of the evolution, these quantities vary slowly, but as the states approach each other, the rigidity starts to vanish, and they start to react strongly to even small variations of the parameters.  At the EP, as shown in Eq.~(\ref{equa}), these quantities diverge. While these divergences yield seemingly unphysical real values of the observables, the accompanying large imaginary components reflect the strong dispersion and intrinsic uncertainty of the system near coalescence, restoring a consistent physical interpretation of the results. 
When we take the sum of the results for both states, the divergences compensate each other, expressing the fact that, although the individual eigenstates are ill-defined at the EP, the total subspace they span remains physically meaningful and observables associated with the combined system remain finite and continuous across the EP, albeit with a slight variation along the evolution as the Hamiltonian is being modified by the control parameters.

\section{Conclusions}
We have investigated the occurrence and impact of EPs in the spectrum of $^{7}$Be using the GSM-CC. By varying the proton and neutron spin-orbit terms of the one-body potential, an EP associated with the $5/2^-$ doublet was identified. Approaching this point, the two states coalesce in energy, total width, as well as in partial widths. The phase rigidity collapses to zero, and the states become strongly mixed and self-orthogonal at the EP. The effects of the latter can be clearly seen in the spectroscopic factors and in the $B(E2)$  values of the separate states. When approaching the EP, they strongly diverge due the self-orthogonality; leading to unphysical real values as the rigidity decreases, while attaining progressively greater imaginary parts implying that the standard deviation of any property which is represented by operators that do not commute with the Hamiltonian would be extremely large under these conditions. Nevertheless, their sum evolves smoothly and is almost constant, showing the non-separability of these states due to their coupling through the environment of continuum states.

\vskip 0.3truecm
\textit{Acknowledgments --}
N. Michel and S. M. Wang wishes to thank GANIL for the hospitality where this work has been done. We thank J.P. Linares Fernandez for useful discussions.
This material is based upon work supported by
the National Natural Science Foundation of China under Grant No.12175281, and the State Key Laboratory of Nuclear Physics and Technology, Peking University under Grant No. NPT2020KFY13, and the National Key Research and Development Program of China (MOST 2023YFA1606404 and MOST 2022YFA1602303); the National Natural Science Foundation of China under Contract No.\,12347106, No.\,12147101, No.12447122, and the China Postdoctoral Science Foundation under Contract No.\,2024M760489.
We gratefully acknowledge support from the CNRS/IN2P3 Computing Center (Lyon, France) and the CRIANN (Normandy, France) for providing computing and data-processing resources needed for this work.

\bibliography{bib}
\bibliographystyle{apsrev4-1}

\end{document}